\documentclass[12pt]{article}
\usepackage{pdproc,epsfig} 

\begin{document}

\renewcommand{\thefootnote}{\alph{footnote}}
  
\title{High-energy neutrinos and hard $\gamma$-rays
 \\ in coincidence with Gamma Ray Bursts}

\author{ A. De R\'ujula}

\address{ Theory Division, CERN,
  Address\\
 1211 Geneva 23, Switzerland\\
 {\rm E-mail: alvaro.derujula@cern.ch}}

 \abstract{The observations suggest that $\gamma$-ray bursts (GRBs) 
are produced by jets of relativistic cannonballs (CBs),
emitted in supernova (SN) explosions. The CBs, reheated by their
collision with the SN shell, emit radiation and Doppler-boost it to
the few-hundred keV energy of the GRB's photons.
Chaperoning the GRB, there should be an intense flux of 
neutrinos of a few hundreds of GeV energy, made in $\pi^\pm$ decays: 
the SN shell acts as a dump of the beam of CBs. The $\nu$ beam
carries almost all of the emitted energy, but
is much narrower than the GRB beam and should only be
detected in coincidence with the small fraction of GRBs whose
CBs are very precisely pointing to us. The $\pi^0$s made in the 
transparent outskirts of the SN shell decay into energetic 
$\gamma$-rays (EGRs) of energy of ${\cal{O}}$(100) GeV.
The EGR beam, whose energy fluence is comparable to that of the 
companion GRB, is
as wide as the GRB beam and should be observable, in coincidence
with GRBs, with existing or planned detectors.}
   
\normalsize\baselineskip=15pt

\section{Introduction}
  
For over thirty years, gamma ray bursts (GRBs) have
been a great astrophysical mystery.
Their origin is an unresolved enigma, in
spite of recent remarkable observational progress: the discovery of 
GRB afterglows\cite{Costa}\cite{vP}
the discovery\cite{Galama} of the association of GRBs with
supernovae (SNe), and the measurements of the redshifts\cite{Metz}
of their host galaxies. The current generally accepted view  is
that GRBs are due to synchrotron
emission from {\it fireballs} produced by collapses
or mergers of compact stars\cite{PG}, by failed supernovae
or collapsars\cite{WM}, or by hypernova explosions\cite{Pacz}.
But various observations suggest that most GRBs are produced by
highly collimated superluminal jets\cite{SD}\cite{Dar98}\cite{DD2000a}.

In a recent series of 
papers\cite{DD2000a}\cite{DD2000b}\cite{DD2001}\cite{DD2001b}
Arnon Dar and I have outlined a cannonball (CB)
model of GRBs which, we contend, is capable of describing the GRB
phenomenology, and results in interesting predictions.
For the sake of fairness and, speaking to an audience of particle
physicists, I must admit that the reception of our CB model in the
astrophysics community has, with a few significant exceptions,
ranged from utter indiference
to militant opposition. Arnon and I, who are optimists, consider this
a very good omen, since in the ever evolving field of GRBs, 
and prior to every major shift in paradigm, the majority has always been
wrong.

\subsection{Interlude on the CB model}

The CB model is based on the following analogies, 
hypothesis and explicit calculations:

{\it Jets in astrophysics.}
Astrophysical systems, such as quasars and microquasars,
in which periods of intense accretion into a massive object occur, emit
highly collimated jets of plasma. The Lorentz factor 
$\rm \gamma\equiv 1/\sqrt{1-v^2/c^2}$ of these jets ranges from 
mildly relativistic: $\gamma\sim 2.55$ for PSR 1915+13\cite{MR}, to
quite relativistic: $\gamma={\cal{O}}\,(10)$ for typical quasars\cite{Ghis},
and even to highly relativistic: $\gamma\sim 10^3$
for PKS 0405$-$385\cite{Ked}. 
These jets are not continuous streams of matter, but consist
of individual blobs, or {\bf ``cannonballs''}. The jets emitted by quasars
must be seen to be believed: they extend for many times the size
of a galaxy, and they are unresolvably pencil-collimated
till their material finally loses its kinetic energy to the intergalactic medium,
stops and expands. The mechanism producing
these surprisingly energetic and collimated emissions is not
understood, but it seems to operate pervasively in nature
(the {\it mantra} here is {\it MHD}, for magneto-hydrodynamics,
which is not yet solved, particularly in its relativistic or
general--relativistic versions.)
We assume the CBs to be composed of ordinary ``baryonic''
matter (as opposed to $\rm e^+\, e^-$ pairs), as is the case in the microaquasar
SS 433, from which Ly$_\alpha$ and metal K$_\alpha$ lines have been detected
\cite{Marg}\cite{Kot}.

\begin{figure}[t]
\begin{tabular}{cc}
\hskip 2truecm
\vspace*{2cm}
\hspace*{-1.5cm}
\epsfig{file=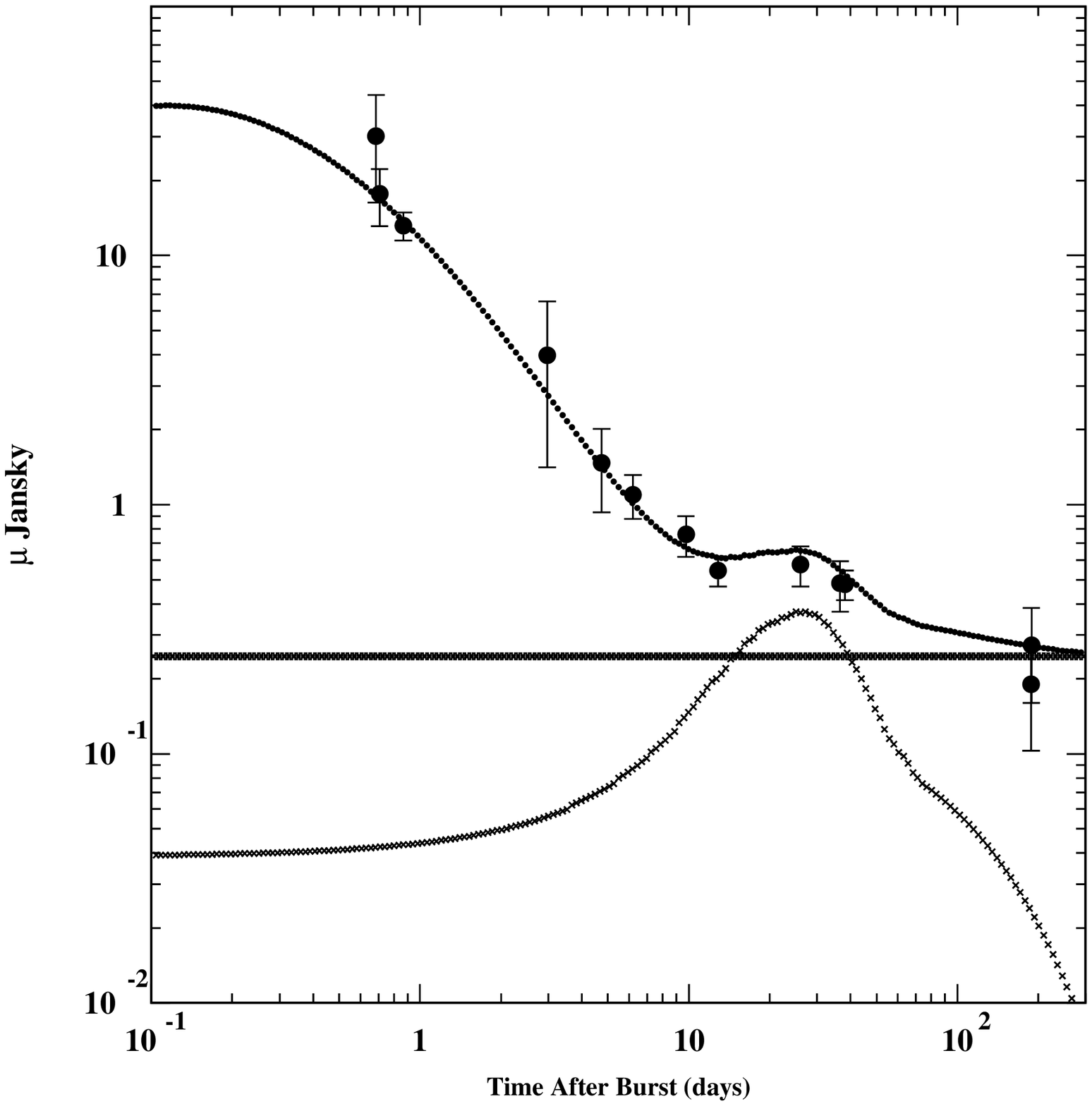,width=10cm} 
\vspace{-2.5cm}\\
\hspace*{.4cm}
\epsfig{file=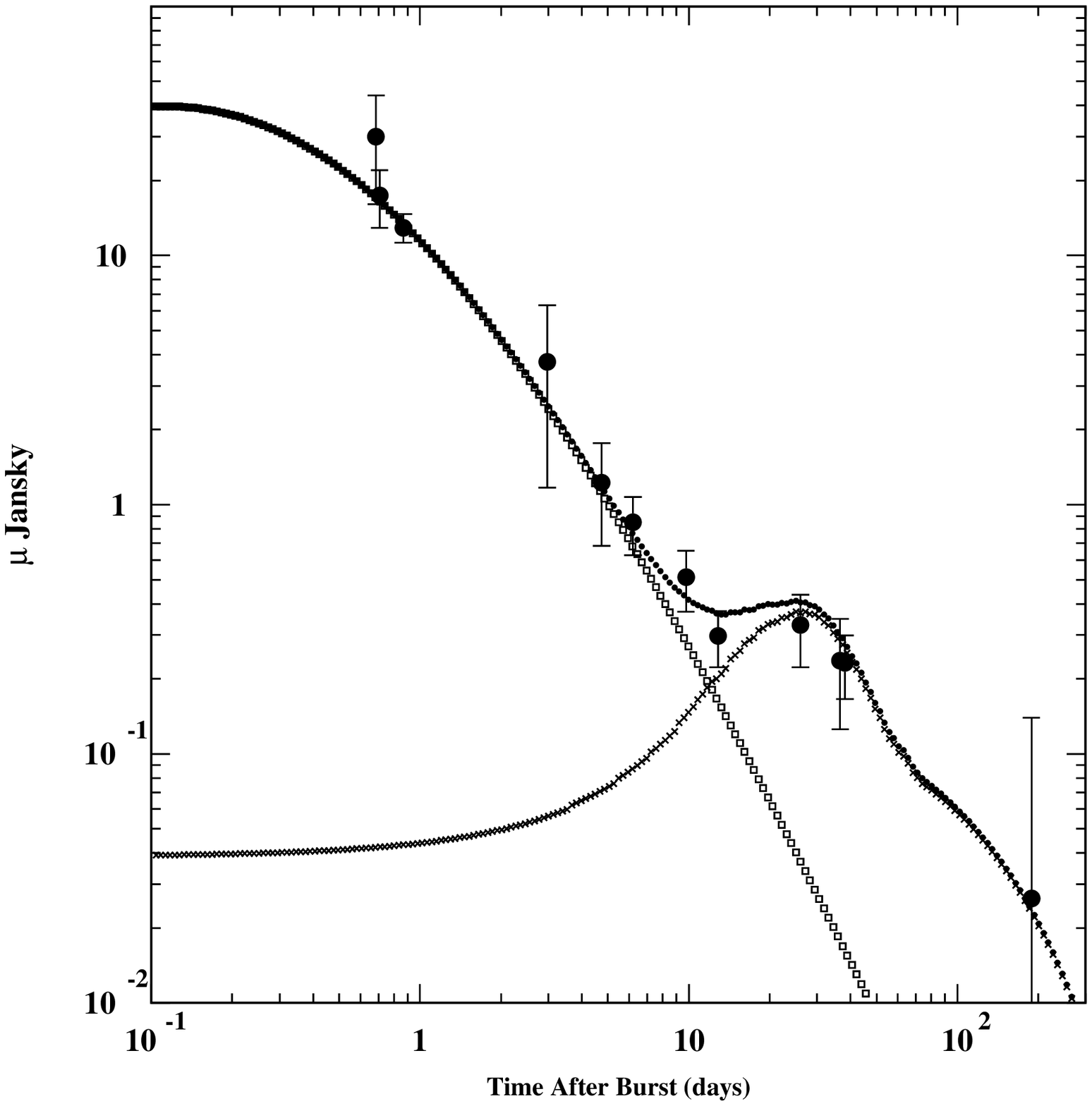,width=10cm}
\end{tabular}
\caption{The R-band afterglow of GRB 970228, as analized in Ref.[20]. 
The upper figure shows the data, a constant background galaxy and
the z-transported 1998bw-like SN-contribution. In the lower
panel the galaxy is subtracted.}
\label{228}
\end{figure}

\begin{figure}[t]
\begin{tabular}{cc}
\hskip 2truecm
\vspace*{2cm}
\hspace*{-0.6cm}
\epsfig{file=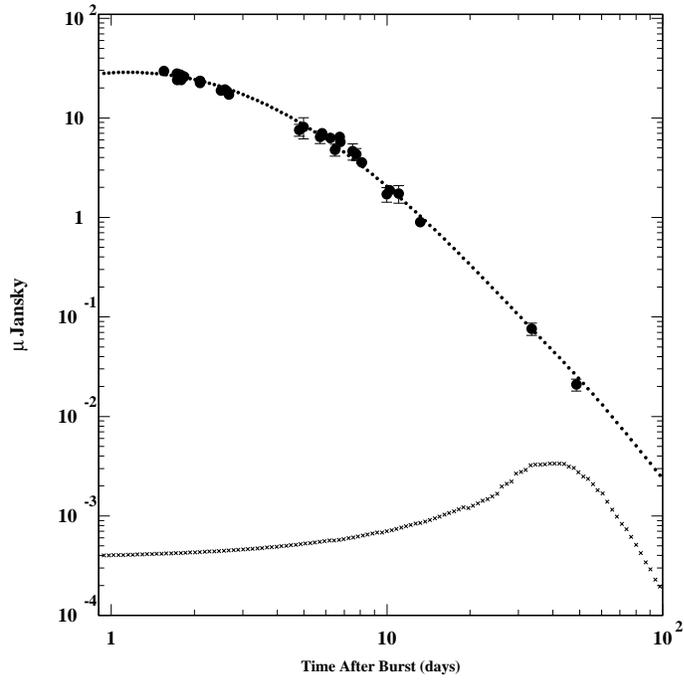,width=10cm} 
\vspace{-2.5cm}\\
\hskip 1truecm
\hspace*{.4cm}
\epsfig{file=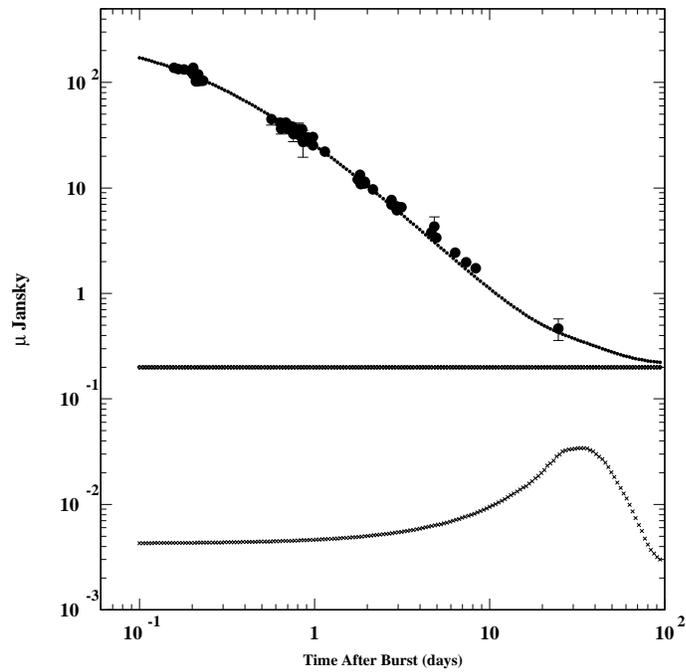,width=10cm}
\end{tabular}
\caption{The R-band afterglows of GRB 000301 and GRB 010222,
as analized in Ref.[20]. 
In the upper figure (000301), the afterglow outshines the 
1998bw-like SN,
and the galaxy is negligible. In the lower one (010222) the galaxy
outshines the SN.}
\label{301}
\end{figure}

{\it The GRB/SN association.}
The original observation of an association between the exceptionally
close-by GRB 980425 (redshift $\rm z=0.0085$) with the supernova
SN 1998bw has developed into a convincing case for the 
claim\cite{DD2000a} that {\it many,  perhaps all, of the long-duration GRBs are
associated with SNe}. Indeed, of the dozen and a half GRBs whose redshift is
known, six or seven show in their afterglow a more or less
significant additive ``bump'', with the time dependence
and spectrum of a SN akin to 1998bw\cite{DD2000a},
properly corrected\cite{Darz} for the different
redshift values. The example\cite{DD2000a} of 
GRB 970228 is given in Fig.(\ref{228}).
In all the other cases\cite{DD2001b} there is a good reason for
such a bump not to have been seen, e.g. the afterglow itself is too luminous,
as in the GRB 000301 example of Fig.(\ref{301}); or the
underlying galaxy is too luminous, as in the
GRB 010222 example in the same figure,
or the observations at the time the SN ought to be prominent are not available:
the best reason not to have seen something\footnote{In 
our work in progress\cite{DDD} we are refining 
the CB model of afterglows, and we have excellent fits to all observed
optical afterglows of known redshift.}.
Thus, observationally, more than six out of sixteen ---and perhaps
all--- of the GRBs of known redshift have a SN associated with them.
The energy supply in a SN event similar to SN 1998bw is too small
to accommodate the fluence of cosmological GRBs, unless their
$\gamma$-rays are highly beamed. SN 1998bw is a peculiar supernova,
but that may be due to its being observed close to the axis of its
GRB emission. {\it It is not out of the question that a good fraction
--perhaps all-- of the core-collapse SNe be associated with GRBs.}
To make the total cosmic rate of GRBs and SN compatible, this
nearly one-to-one GRB/SN association would require beaming into
a solid angle that is a fraction $\rm f\sim 2\times 10^{-6}$ of $4\pi$
\cite{DD2000a}. The CB model, for the emission from CBs moving
with $\gamma\sim 10^3$, implies precisely that beaming factor.

\begin{figure}
\begin{center}
\vspace*{1.0cm}
\hspace*{-1cm}
\epsfig{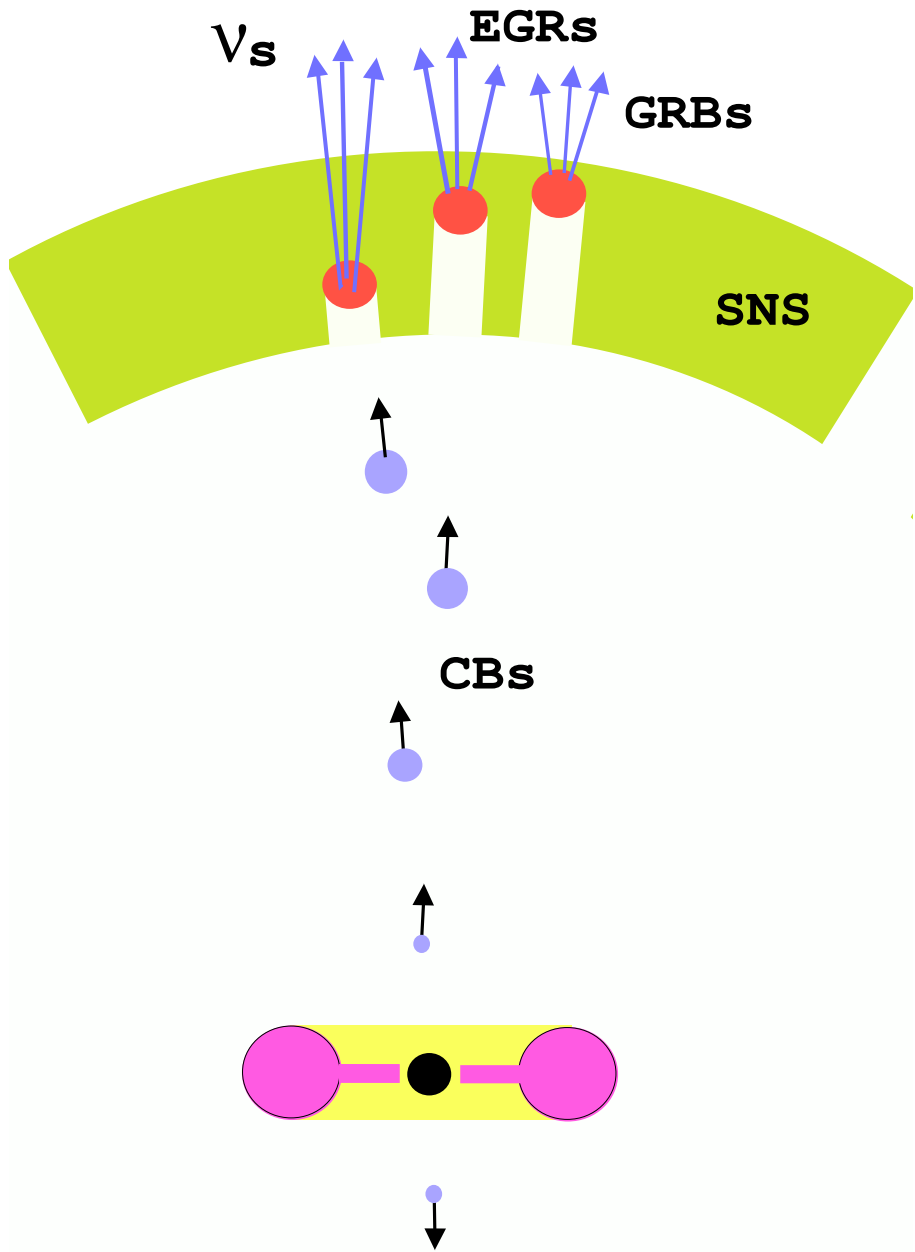}
\caption{The CB model in a SN environment, not shown to scale.
Relativistic CBs are emitted by a compact object
accreting matter from a disk and/or torus.
They hit a SN shell generating $\nu$'s,
quasi-thermal radiation
(the GRB) and $\gamma$-rays from $\pi^0$ decay
(the EGRs). The latter two exit only from the
transparent outer layers of the SN shell.}
\vspace*{-0.5cm}
\label{model}
\end{center}  
\end{figure}

{\it The GRB engine.}
We assume a core-collapse SN event not to result only in the
formation of a central compact object and the expulsion of
a supernova shell (SNS). A fraction of the parent star's
material, external to the newly-born compact object,
should fall back in a time very roughly of the
order of one day\cite{DD2000a} and, given the considerable
specific angular momentum of stars,
it should settle into an accretion disk and/or torus
around the compact object.
The subsequent sudden episodes
of accretion ---occurring with a time sequence that we cannot predict---
result in the emission of CBs. These emissions last till
the reservoir of accreting matter is exhausted.
The emitted CBs initially expand in the SN rest-system at a speed 
$\rm\beta\,c/\gamma$, with $\rm\beta\,c$ presumably of the same order or 
smaller than the speed of sound in a relativistic plasma ($\beta=1/\sqrt{3}$).
The solid angle a CB subtends is so small that presumably
successive CBs do not hit the same point of the outgoing SNS,
as they catch up with it. These considerations 
are illustrated in Fig.(\ref{model}).

{\it The GRB.}
From this point onwards, the CB model is not based on analogies or
assumptions, but on processes whose outcome can be approximately 
worked out in an explicit manner. The violent collision of the CB
with the SNS heats the CB (which is not transparent at this point
to $\gamma$'s from $\pi^0$ decays) to a temperature that, by
the time the CB reaches the transparent
outskirts of the SNS, is $\sim 150$ eV,
further decreasing as the CB travels\cite{DD2000b}.
The resulting CB surface radiation, Doppler-shifted in energy 
and forward-collimated by the CB's fast motion, gives rise to an individual
pulse in a GRB, as illustrated in Fig.(\ref{model}).
The GRB light curve is an ensemble of such pulses,
often overlapping one another.
The energies of the individual GRB $\gamma$-rays, as well as
their typical total fluences, indicate CB Lorentz factors of 
${\cal{O}}$(10$^3$), as the SN/GRB association does\cite{DD2000b}.
The observed fluence, light curves and energy spectra of GRBs
are well described by the CB model\cite{DD2000b}.

{\it The GRB's afterglow.}  The CBs, after they exit the SNS,
cool down by brems-strahlung and radiate by this process,
by inverse Compton scattering, and by 
synchrotron radiation of their electrons on their enclosed magnetic field,
much as the plasmoids emitted by quasars and microquasars 
do\cite{DD2000a}.  The CB model provides an excellent detailed 
description of optical afterglows\cite{DDD}.
The early afterglow spectrum and light curve are complicated
by the fact that, about a day after the GRB emission, CBs cool down to
a temperature at which $\rm e$--$\rm p$ recombination into $\rm H$
takes place. This gives rise to Ly-$\alpha$ lines that the CB's motion
Doppler-shifts to (cosmologically redshifted) energies of order 
a few keV, an energy domain that, interestingly,
coincides with that of the $\rm Fe$ lines that an object at rest would emit.
Recombination also gives rise to a multiband-flare in the afterglow.
These CB-model's expectations are in good agreement with incipient
data on X-ray lines and flares\cite{DD2001}, but the data are at the moment
too scarce and imprecise to offer a decisive test of either the CB model
or the Fe-line interpretation.

\subsection{Back to introductory remarks}

In this talk I concentrate on the $\nu$ and EGR signals that
should occur in directional and temporal coincidence
with GRBs.
 The neutrinos are made by the chain decays of
charged pions, produced in the collisions of the CBs'
baryons with those of the SNS, as in Fig.(\ref{model}). 
The $\nu$ beam carries almost all of the emitted energy, but
is much narrower than the GRB beam and should only be
detected in coincidence with the small fraction of GRBs whose
CBs are moving extremely close to the line of sight.
The EGRs are made by the decay of neutral pions, but only from $\pi^0$
production close enough to the outskirts of the SNS for the $\gamma$-rays
not to be subsequently absorbed, see Fig.(\ref{model}).
The EGR beam, whose fluence is comparable to that of the GRB, is
as wide as the GRB beam and should be observable, in coincidence
with GRBs, with existing or planned detectors.
The EGR beam peaks at energies of tens of GeVs, while the
$\nu$ beam is about one order of magnitude more energetic.

In the space of a conference's proceedings, I cannot  detail
the derivation of these results. In this written version I am mostly
stating results, often in the form of commented figures.

\section{Times and energies}

To understand the CB model, one must keep four clocks
simultaneously ticking on one's head.
Let $\rm \gamma=1/\sqrt{1-\beta^2}={E_{CB}/(M_{CB}c^2)}$ be 
the Lorentz factor
of a CB, which diminishes with time as the CB hits the SNS
and as it subsequently plows through the interstellar medium. 
Let $\rm t_{SN}$ be the
local time in the SN rest system, $\rm t_{CB}$ the time in the CB's 
rest system, $\rm t_{Ob}$ the time   measured by
a nearby observer viewing the CB at an angle $\theta$
away from its direction of motion, and $\rm t$ the time
measured by an earthly observer viewing the CB at
the same angle, but from a ``cosmological'' distance 
(redshift $\rm z\neq 0$).
Let x be the distance traveled by the CB in the SN rest system.
The relations between the above quantities are:
\begin{eqnarray}
&&\rm
dt_{SN}=\gamma\,dt_{CB}=\rm{dx\over\beta\, c}\, ;
\nonumber \\
&&\rm
dt_{CB}\equiv \delta\,dt_{Ob}\, ;\nonumber\\ 
&&\rm
dt=(1+z)\,dt_{Ob}={1+z\over \gamma\,\delta}\;dt_{SN}\;,
\label{times}
\end{eqnarray}
where the Doppler factor $\delta$ is:
\begin{equation}
\rm
\delta\equiv\rm{1\over\gamma\,(1-\beta\cos\theta)}
\simeq\rm {2\,\gamma\over (1+\theta^2\gamma^2)}\; , 
\label{doppler} 
\end{equation}
and its approximate expression is valid for $\theta\ll 1$ and $\gamma\gg 1$,
the domain of interest.

Notice that for large $\gamma$ and not large $\theta\gamma$,
there is an enormous ``relativistic aberration'':
$\rm dt\sim dt_{SN}/\gamma^2$, and the observer sees
a long CB story as a film in extremely fast motion. As an example,
consider a feature of a GRB light curve, such as a pulse
of 2 s duration. Assume it is due to a GRB at $\rm z=1$,
for which $\gamma=10^3$, and the viewing angle is 
$\theta=1/\gamma$. The duration of the pulse in the vicinity of
the SN would be 1 s, corresponding to $10^3$ seconds in
the CB system, and $10^6$ s (11.6 days) of travel of the CB
through interstellar space. Everything that happens to the CB in
its journey of a dozen light-days, the observer sees in 2 seconds!
 
The energy of the photons radiated by a CB
in its rest system, $\rm E^\gamma_{CB}$, their energy
in the direction $\theta$
in the local SN system, $\rm E^\gamma_{SN}$,  and the photon
energy $\rm E$ measured by a cosmologically distant observer,
are related by:
\begin{equation}
\rm E^\gamma_{CB}=   {E^\gamma_{SN}\over \delta}
\, ;\;\;\;\;\;E^\gamma_{SN}=(1+z)\,E\; ,
\label{energies}
\end{equation}
with $\delta$ as in Eq.(\ref{doppler}). Relative to their energy
at the SN location, the GRB's photons are blue-shifted by
a factor $\delta$ and red-shifted by $\rm 1+z$. This brings
the photon energy of the CB's surface radiation to the observed
GRB domain.

\section{Reference values of various parameters}

To be explicit we must scale our results to given values of the 
parameters of the CB model, which serve as benchmarks 
but imply no strong commitment to their particular choices.
These values are listed in Table I, for quick reference.

\begin{table}[h]
\hspace{2.5 cm}
\begin{tabular}{|l|c|c|c|}
\hline
\hline
$\;\;\;\;\;\;\;\;\;$Parameter   &Symbol &Value \\
\hline
SN-shell's mass     & $\rm M_S$             & $\rm 10\; M_\odot$ \\
SN-shell's radius   & $\rm R_S$              & $2.6\times 10^{14}$ cm \\
Outgoing Lorentz factor   & $\rm\gamma_{out}$  & $10^3$ \\
CB's energy   & $\rm E_{CB}$ & $10^{52}$ erg   \\
Initial $\rm v_{_T}/c$ of expansion & $\rm\beta_{in}$ & $1/(3\,\sqrt{3})$ \\
Final $\rm v_{_T}/c$ of expansion & $\rm\beta_{out}$ & $1/\sqrt{3}$ \\
\hline
Redshift   &z  &1   \\
CB's viewing angle  & $\theta$ & $\rm 10^{-3}$  \\
\hline
\hline
\end{tabular}
\caption{List of the ``reference'' values of various parameters. 
$\rm z$ and $\theta$ 
are not specific to the model.}
\end{table}

Let ``jet'' stand for the ensemble of CBs emitted in one direction in a SN
event. If a momentum imbalance between the opposite-direction jets is
responsible for the large peculiar velocities of neutron stars\cite{LL},
${\rm v_{NS}\approx 450\pm 90~ km~s^{-1}}$, the
jet kinetic energy $\rm E_{jet}$ must be, as we shall assume for our GRB
engine, larger than\cite{DP} $\rm M_{NS}\,v_{NS}\,c\sim 10^{52}$ erg,
for $\rm M_{NS}=1.4\,M_\odot$. 
We adopt a  value of $10^{53}$ ergs as the reference 
jet energy. 
On average, GRBs have some five to ten significant pulses, so that the 
energy in a single CB may be 1/5 or 1/10 of $\rm E_{jet}$. We adopt 
$\rm E_{CB}=10^{52}$ erg as our reference.

Let $\rm \gamma_{in}$ be the Lorentz factor of a cannonball
as it is fired. We find $\rm\gamma_{in}={\cal{O}}(3\times 10^3)$
to be a ``typical'' value ($\rm\gamma_{in}$ is not an ``input'' parameter).
For this value and the reference CB energy, the CB's mass is
very small by stellar standards, and comparable to an Earth mass:
$\rm M_{CB}\sim 0.6\, M_\otimes \;{(3\times 10^3)/\gamma_{in}}$.
The baryonic number of the CB is:
\begin{equation}
\rm N_b\simeq {E_{CB}\over m_p\,c^2\,\gamma_{in}}\simeq 2.2\times 10^{51}\; 
{{E}_{CB}\over 10^{52}\; erg}
\, \left[{3\times 10^3\over\gamma_{in}}\right].
\label{NB}
\end{equation}
Thus, the CB accelerator is comparable to the Tevatron in energy per proton
($\rm E_p={\cal{O}}(1)$ TeV), but it has a better flux ($10^{51}$ p.p.p.)
The trouble is that nobody has a convincing idea about how the 
accelerating trick is played.

We have assumed that, in a SN explosion, some of the material
outside the collapsing core is not expelled
as a SNS, but falls back onto the compact object. For vanishing angular momentum, the free-fall time of a test-particle from a distance $\rm R$
onto an object of mass ${\rm M}$ is
$\rm t_{fall} =\pi\,[R^3/(8\,G\,M)]^{1/2}$. For material falling
from a typical star radius ($\rm R_\star\sim 10^{12}$ cm) on an
object of mass $\rm M=1.4\;M_\odot$, $\rm t_{fall} \simeq 1$ day.
The fall-time is longer (except for material falling from the polar directions)
if the specific angular momentum is considerably
large, as it is in most stars. The fall-time is shorter for material
not falling from as far as the star's radius.
The estimate $\rm t_{fall} \simeq 1$ day is therefore a very rough one.
One day after core-collapse, the expelled SNS, travelling at
a velocity\cite{Naka} $\rm v_S \sim c/10$,
has moved to a distance of $\rm R_S=2.6 \times 10^{14}$ cm,
which we adopt as our reference value.

For the Lorentz factor of the CBs as they exit the SNS, 
we adopt the value $\rm \gamma_{out}=10^3$, 
for the reasons discussed in the Introduction.
Let $\rm \beta_{in}\, c$ be the expansion velocity of a CB,
in its rest system, as it travels from the point of emission to
the point at which it reaches the SNS, and let $\rm \beta_{out}\, c$
be the corresponding value after the CB exits the SNS,
reheated by the collision. We expect these velocities to be comparable 
to the speed of sound in a relativistic plasma, $\rm c/\sqrt{3}$,
as observed in the initial expansion of the CBs
emitted by GRS 1915+13\cite{MR}. As reference values, we
adopt those of Table I.

\section{A cannonball's collision with a supernova shell}

The density profile of the material in a SNS 
(as a function of distance $\rm x$ to the SN centre) is observationally known,
at least for the outer transparent regions of the shell\cite{Naka}. 
It is roughly given as a power law $\rm \rho\propto x^{-n}$,
with $\rm n=4$ to 8. Given the assumed properties of our CBs,
we can approximately figure out what happens in their collision
with a SNS of given mass, radius and density index $\rm n$.

The CB and the SNS are sufficiently thin for the charged pions
produced in $\rm pp$ (or nucleon-nucleon)
collisions and the subsequent muons to
decay: about two thirds of the original energy of a colliding
CB's nucleon results in $\nu$ production.  The CB and the
SNS are sufficiently thick for all incoming nucleons to interact.
The CB's energy is much bigger than the ``target'' rest energy
of the material removed from the ``bullet-hole'' which the CB makes
in the SNS. It follows from all this that the relation between the
Lorentz factors of the CB before and after it crosses the SNS, is:
\begin{equation} 
\rm \gamma_{out} \simeq \gamma_{in} 
\;\sqrt{2\,E_{CB}\over 3\,\beta_{in}^2\,M_S\, c^2+18\,E_{CB}}\, .
\label{gammaout}
\end{equation}
For large $\rm E_{CB}$, $\rm \gamma_{in}\sim 3\, \gamma_{out}$,
while for our reference parameter values 
$\rm \gamma_{in}\sim 10\, \gamma_{out}$: the model requires
values of $\rm \gamma_{in}$ as surprisingly large as $10^4$.

After the CB and the target SNS material collide and coalesce,
all nucleons of the CB must have slowed down from $\rm\gamma_{in}$
to $\rm\gamma_{out}$, which takes a few collisions of high 
c.m.s.~energy. The neutrinos made in these collisions escape.
I shall outline how this allows us to estimate the $\nu_\mu$ flux 
as a function of energy and production angle explicitly, by use
of the empirical knowledge on pion and ``leading particle''
production in nucleon-nucleon collisions.

One can similarly estimate the EGR flux.
The CB-SNS collision results in the production of $\pi^0$s.
The energetic $\gamma$ rays from $\pi^0\rightarrow \gamma\gamma$ decay,
of energy a few hundred GeV,
can only escape from the transparent outskirts of the SNS, of grammage
$\rm X_{EGR}\sim 70$ g/cm$^2$: their observable flux is much smaller
than that of neutrinos. The number of ``target'' SNS nucleons involved in 
the production of escaping EGRs is:
\begin{equation}
\rm N_p^{EGR} \approx  \pi\;\widetilde R_{CB}^2\,{X_{EGR}\over m_p}
\simeq 1.4\times 10^{49}\; ,
\label{Np}
\end{equation}
where $\rm \widetilde R_{CB}$ the radius of the CB as it reaches the
transparent outer shell. The numerical value is for
our reference parameters (the general expression\cite{DD2001b}
for $\rm N_p^{EGR}$ is complicated).

\begin{figure}
\begin{center}
\vspace*{1.0cm}
\hspace*{-1cm}
\epsfig{file=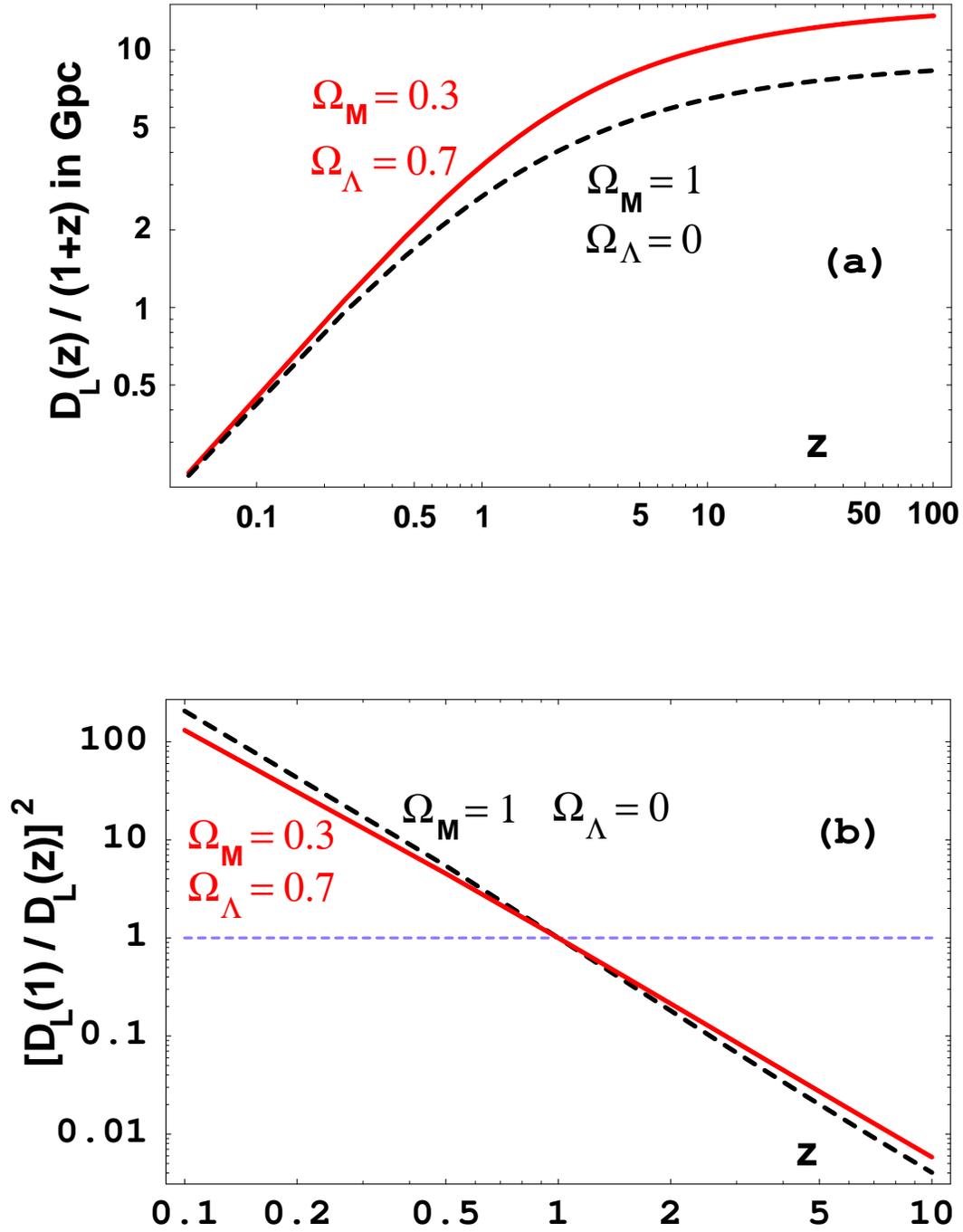,width=14cm}
\caption{Luminosity distances and ratios thereof, as functions of redshift,
for two $\Omega=1$ Friedman universes, with
two choices of matter and vacuum densities.}
\vspace*{-0.5cm}
\label{lumdis}
\end{center}  
\end{figure}

In the CB model the GRB $\gamma$ rays, of much lower energy
than the EGRs, result from the quasi-thermal emission from the
CB's front-surface, heated by the collision with the SNS. Their
detailed discussion is elaborate\cite{DD2000b}. For our 
reference parameters, the total GRB energy in
the CB rest system is $\rm E_{pulse}^{rest}\sim 3\times 10^{45}$ erg. 
An observer at rest,
located at a known luminosity distance $\rm D_L(z)$ from the CB and
viewing it at an angle $\theta$ from its direction of motion, would measure
a  ``total'' (time- and energy-integrated) fluence per unit area:
\begin{equation}
\rm {df\over d\Omega}\simeq {1+z\over 4\,\pi\,D_L^2}
\,{E_{pulse}^{rest}}\;\left[
{2\,\gamma_{out} \over 1+\theta^2\,\gamma_{out}^2}\right]^3\; .
\label{dfdomega}
\end{equation}
Of the three powers of $\delta$, the Doppler factor of Eq.(\ref{doppler}),
one reflects the energy boost, and two the narrowing of the
(solid) angle. The photons in the
GRB are beamed in a cone of aperture $\rm 1/\gamma_{out}$.
In our explicit calculations, and in the conventional cosmological notation,
we use $\rm H_0=65$ km/(s Mpc), ${\rm \Omega_M}=0.3$ and 
${\rm \Omega_\Lambda}=0.7$, so that, for example, 
$\rm D_L(1)\simeq 7.12$ Gpc $\simeq 2.20\times 10^{28}$ cm. In 
Fig.(\ref{lumdis}) I show $\rm D_L(z)$ and $\rm [D_L(1)/D_L(z)]^2$
(the quantity to which we shall scale our results) for the quoted
cosmology and, for comparison, for the case $\rm\Omega_M=1$,
$\rm \Omega_\Lambda=0$. At moderate $\rm z$,
the results are not very sensitive to the choice of cosmological parameters.

\section{The flux of EGRs}

The approximate isospin independence of nuclear interactions
implies that, in the high energy collisions of protons or neutrons
on protons or neutrons, the production of 
$\pi^+$, $\pi^0$ and $\pi^-$ is similar. 
To compute the spectrum of outgoing photons {\it per nucleon--nucleon
collision}, we convolute the observed
distribution of $\pi^\pm$s (assumed to describe $\pi^0$ production as well)
with the $\gamma$ distribution in $\pi^0$ decay.
The resulting  distribution
in $\rm y\equiv E_\gamma/E_p$ and $\gamma$
transverse momentum $\rm p_T$,
can be fitted by:
\begin{eqnarray}
\rm 
{1\over\sigma_{pp}^{TOT}}\,
{d\sigma^\gamma\over dp_{_T}^2\, dy}&\simeq & \rm
\rm A_\gamma\,{1\over y}
\,e^{-b_\gamma y}\;
{1\over 2\, \bar{p}_T^2}\,e^{-p_{_T}/\bar p_{_T}}\; , \nonumber \\
\rm
A_\gamma&\simeq & \rm 1.1\; ,\;\;\; b_\gamma\simeq 8\,, \;\;\;\;
\bar p_{_T}\sim 160~MeV\, .
\label{pTydistr}
\end{eqnarray}
An exponential fit in $\rm y$ is inadequate close
to the limit $\rm y=1$, but for $\rm y>1/2$ the
flux is negligible. The advantage of our simple fits is that they allow
us to give analytical estimates all the way to the observable 
particle fluxes.

Let $\rm E_\gamma$ be the energy of a photon as it
reaches the Earth, cosmologically
red-shifted by a factor $\rm 1+z$, and let
$\rm E_p\simeq m_p\,c^2\,\gamma_{out}$ be the energy 
of the CB's nucleons,
 in the local rest system of their SN progenitor,
as they reach the outer part of the SNS. For the
small angles $\theta$ at which the $\gamma$-rays are forward-collimated
by the relativistic motion of the parent $\pi^0$'s, the photon-number
distribution in $\rm x_\gamma=E_\gamma/E_p$
and $\cos\theta$, per single nucleon--nucleon
collision, is:
\begin{eqnarray}
\rm
{dn_\gamma\over dx_\gamma\;d\cos\theta}&\simeq & \rm
B_\gamma\; (1+z)^2\;x_\gamma\; e^{-c_\gamma\,x_\gamma}\nonumber\\ \rm
B_\gamma&\simeq&\rm A_\gamma\;
\left[{m_p\,c\,\gamma_{out}\over \bar p_{_T}^\gamma}\right]^2
\simeq (3.76\times 10^7)\,\left[{\gamma_{out}\over 10^3}\right]^2\nonumber\\ \rm
c_\gamma&= &\rm c_\gamma (z,\theta,\gamma_{out})\simeq \rm ( 1+z)\,
\left[b_\gamma+{m_p\,c\,\gamma_{out}\,\theta\over\bar p_{_T}^\gamma}\right]\, .
\label{xtheta1}
\end{eqnarray}

Let $\rm dn_\gamma/d\Omega$ be the total (time-integrated)
number flux of EGR photons
per unit solid angle about the direction $\theta$ (relative to
the CBs' direction of motion)  at which they
are viewed from Earth. 
The photon number distribution  per incident CB is:
\begin{eqnarray}
\rm 
{dn_\gamma\over dx_\gamma\, d\Omega}&\sim& \rm
{N_p^{EGR}\,B_\gamma\over 2\,\pi\;D_L^2}\;(1+z)^4\;f_\gamma\nonumber\\
\rm f_\gamma&\equiv&\rm f_\gamma (z,\gamma_{out},\theta,x_\gamma)
\simeq
x_\gamma\; e^{-c_\gamma\,x_\gamma}\; ,
\label{photnum}
\end{eqnarray}
with $\rm N_p^{EGR}$ given by Eq.(\ref{Np}). Since a
typical GRB has an average of $\rm n_{CB}=5$ to 10 significant pulses,
the total flux of EGRs in coincidence with a GRB may be
an order of magnitude above that of Eq.(\ref{photnum}).
In Fig.(\ref{spectrax}a) we show $\rm f_\gamma$ as a function of
$\rm x_\gamma$ at various $\theta$;
for $\rm z=1$ and $\rm\gamma_{out}=10^3$. 
The average fractional EGR energy in the spectrum of Eq.(\ref{photnum})
is $\rm\bar x_\gamma=2/c_\gamma$, corresponding, at
$\rm z=1$ and for $\rm\gamma_{out}=10^3$, to average energies
$\rm\bar E_\gamma\sim 120$ GeV
for $\theta=0$, $\rm\bar E_\gamma\sim 70$ GeV for 
$\rm \theta=1/\gamma_{out}$,
and $\rm\bar E_\gamma\sim 40$ GeV for $\rm \theta=3/\gamma_{out}$, 
a more probable angle of detection\cite{DD2000a}.
Except at the highest of these energies and/or at redshifts
well above unity, the absorption of $\gamma$-rays on the infrared
background ---for which we have not 
corrected Eq.(\ref{photnum})--- is negligible.

\begin{figure}
\begin{center}
\vspace*{1.0cm}
\hspace*{-1cm}
\epsfig{file=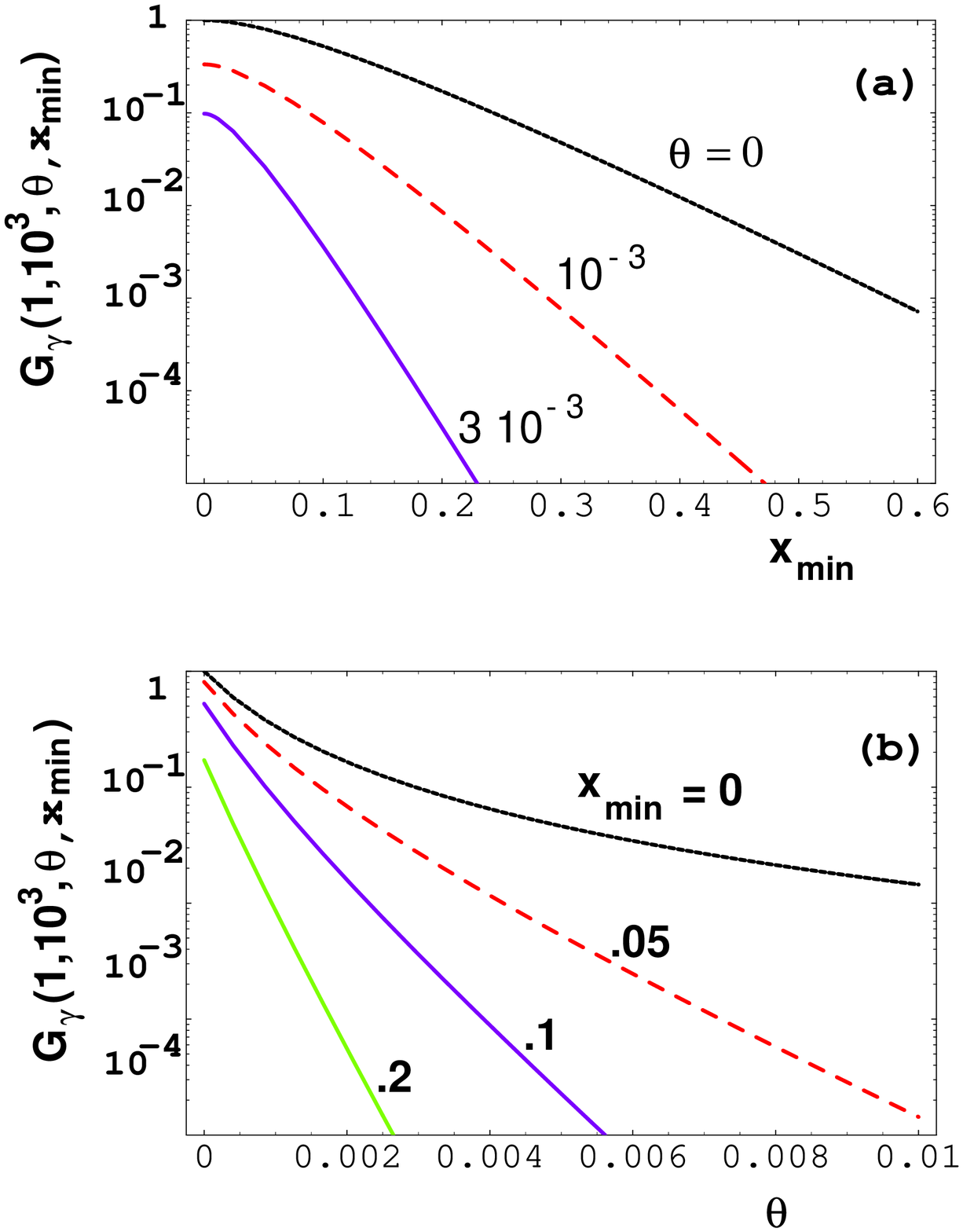,width=14cm}
\caption{The function $\rm G_\gamma$ of Eq.(\ref{photflux2}), for
$\rm z=1$ and $\rm \gamma_{out}=10^3$. Top: As a function of
$\rm x_{min}$ at various fixed $\theta$. Bottom: vice versa.}
\vspace*{-0.5cm}
\label{Ggamma}
\end{center}  
\end{figure}

Roughly characterize the efficiency of a $\gamma$-ray detector
as a step function $\rm \Theta(E^\gamma-E^\gamma_{min})$.
The total flux above threshold, per incident CB, is then:
\begin{eqnarray}
\rm {dn^T_\gamma[x^\gamma_{min},\theta]\over d\Omega}&\sim&\rm
{dn^T_\gamma[0,0]\over d\Omega}\;
G_\gamma (z,\gamma_{out},\theta,x^\gamma_{min})\nonumber\\ \rm   
G_\gamma&\simeq&\rm \left[{(1+z)\,b_\gamma\over c_\gamma}\right]^2
\,(1+c_\gamma\,x^\gamma_{min})\,
e^{-c_\gamma\,x^\gamma_{min}}\nonumber\\ \rm
x^\gamma_{min}&\equiv&\rm {E^\gamma_{min}\over m_p\,\gamma_{out}}
\nonumber\\ \rm
{dn^T_\gamma[0,0]\over d\Omega}
&\simeq&\rm {1.1\times 10^8\over km^2}\,{N_p^{EGR}\over 1.4\;10^{49}}\,
\left[{\gamma_{out}\over 10^3}\right]^2  
\left[{1+z\over 2}\right]^2\,\left[{D_L(1)\over D_L(z)}\right]^2\, .
\label{photflux2}
\end{eqnarray}
In Fig.(\ref{Ggamma}a) and (\ref{Ggamma}b) 
we show $\rm G_\gamma$ as a function of
$\rm x^\gamma_{min}$ at various fixed $\theta$, and vice versa;
for $\rm z=1$ and $\rm\gamma_{out}=10^3$. The very large flux
$\rm dn^T_\gamma[0,0]/ d\Omega$ of Eq.(\ref{photflux2}) is seen
to be significantly reduced as soon as $\theta$ and/or $\rm x^\gamma_{min}$
depart from zero: the EGR flux is not as gigantic as it appears to be
at first sight.

\section{The flux of high energy neutrinos}

The calculation of the $\nu_\mu$ flux 
produced in the collision of  a CB with the 
SNS is analogous to the calculation
of the EGR photon flux. The $\bar\nu_\mu$ flux gives rise to a signal
of about 1/3 the size of that of the $\nu_\mu$ flux
(we neglect it, since we find it preferable to establish
a lower limit to the observational prospects).
The $\nu_\mu$'s are made in the chain reactions
$\rm p\,p\to \pi + ...$, $\pi^+ \to \mu^+\,\nu_\mu$; and 
$\pi^-\to\mu^-\,\bar{\nu}_\mu$, followed by 
$\mu^-\to e^-\,\nu_\mu\,\bar{\nu}_e$.
The contribution of $\rm K$ production and decay turns out to be negligible.

The calculation of the $\nu_\mu$ flux is akin to that of the EGR flux.
Absorption in the SNS no longer plays a role, and $\nu$s
are produced by the collisions of all the nucleons in the CB
as they suffer a few high c.m.s.~energy collisions to slow
down from $\rm\gamma_{in}$ to $\rm\gamma_{out}$. 
The result of our analysis is a distribution
in $\rm y=E_\nu/E_p$ (with $\rm E_p$ the incoming nucleon's
energy) and $\rm p_T^\nu$, which is of the same from as 
Eq.(\ref{pTydistr}), but for which
\begin{equation}
\rm
A_\nu\simeq  \rm 3\; ,\;\;\; b_\nu\simeq 12\,,\;\;\;\; 
\bar p_T^\nu\sim 190~MeV\, .
\label{fnu}
\end{equation}

Let $\rm E_\nu$ be the  redshifted energy of a neutrino as it
reaches the Earth, and let
$\rm E_p\simeq m_p\,c^2\,\gamma_{in}$ be the energy 
of the CB's nucleons, in the local rest system of their SN progenitor, 
as they enter the SNS. In analogy with Eq.(\ref{xtheta1}) the $\nu_\mu$-number
distribution in $\rm x_\nu=E_\nu/E_p$
and $\cos\theta$, per single nucleon--nucleon collision, is:
\begin{eqnarray}
\rm
{dn_\nu\over dx_\nu\;d\cos\theta}&\simeq & \rm
B_\nu\; (1+z)^2\;x_\nu\; e^{-c_\nu\,x_\nu}\nonumber\\ \rm
B_\nu&\simeq&\rm A_\nu\;
\left[{m_p\,c\,\gamma_{in}\over \bar p_{_T}^\nu}\right]^2\simeq
(6.0\times 10^7)\,\left[{\gamma_{in}\over 10^4}\right]^2\nonumber\\ \rm
c_\nu&= & \rm c_\nu (z,\theta,\gamma_{in})
\simeq  ( 1+z)\,
\left[b_\nu+{m_p\,c\,\gamma_{in}\,\theta\over\bar p_{_T}^\nu}\right]\, .
\label{xtheta3}
\end{eqnarray}

Notice that the above equations contain $\rm\gamma_{in}$,
and not $\rm\gamma_{out}$, as the analogous ones for EGRs did.
If you have not understood this, I have lost you already.
Let $\rm dn_\nu/d\Omega$ be the time-integrated number of neutrinos
per unit solid angle about the direction $\theta$ (relative to
the CBs' direction of motion)  at which they
are viewed from Earth. In analogy with Eq.(\ref{photnum}),
the neutrino number distribution, per incident CB, is:
\begin{eqnarray}
\rm 
{dn_\nu\over dx_\nu\, d\Omega}&=& \rm
{N_b\,B_\nu\over 2\,\pi\;D_L^2}\;(1+z)^4\;f_\nu\nonumber\\ \rm
f_\nu&=&\rm f_\nu(z,\gamma_{in},\theta,x_\nu)\simeq
x_\nu\; e^{-c_\nu\,x_\nu}\; ,
\label{nunum}
\end{eqnarray}
with $\rm N_b$ the CB's baryon number,
given by Eq.(\ref{NB}). For a GRB with $\rm n_{CB}$ significant pulses,
the total number of neutrinos is $\rm n_{CB}$ times
larger than that of  Eq.(\ref{nunum}).

In Fig.(\ref{spectrax}b) we show $\rm f_\nu$ as a function of
$\rm x_\nu$ at various $\theta$;
for $\rm z=1$ and $\rm\gamma_{in}=10^4$. 
The average fractional $\nu$ energy in the spectrum of Eq.(\ref{nunum})
is $\rm\bar x_\nu=2/c_\nu$, corresponding, for the chosen $\rm z$
and $\rm\gamma_{in}$, to average energies
$\rm\bar E_\nu\sim 712$ GeV
for $\theta=0$, $\rm\bar E_\nu\sim 315$ GeV for 
$\rm \theta=1/10^3$,
and $\rm\bar E_\nu\sim 150$ GeV for $\rm \theta=3/10^3$.

Neutrino oscillations may reduce the flux of $\nu_\mu$s of
Eq.(\ref{nunum}) by as much as a factor of 2 (if they are
maximal) or even 3 (if they are ``bimaximal'').

\subsection{Muon production on Earth}

Muon neutrinos produced by a GRB can be detected by large-area
or large-volume detectors, in temporal and directional
coincidence with a GRB $\gamma$-ray signal. 
The detection technique typically involves
the ``upward-going'' muons, for which there is no ``atmospheric''
cosmic-ray background. To compute the number of muons of a fixed
energy observable
per unit area, one must solve the quasi-equilibrium equation
describing the competing processes of muon production
and muon energy-loss. At the relatively low energies of interest
here, the details pile up into a single prefactor:
\begin{equation}
\rm
\rm K \simeq  \rm 
\rho_W\,N_A\,{1\over R_0}
\,{\sigma_{_{CC}}\over E_\nu}\simeq 2.26 \times 10^{-12}\;GeV^{-2}\; ,
 \label{K1}
\end{equation}
where $\rm \rho_W$ is the density of water, $\rm N_A$ is
Avogadro's number, $\rm R_0$ is an energy loss of 2.12 MeV/cm
and $\rm \sigma_{CC}/E_\nu\sim 0.8\;10^{-38}$ cm$^2$/GeV is the
slope of the charged current $\nu_\mu$ cross section on an isoscalar nucleon.

Define $\rm x_\mu=E_\mu/E_p$: the ratio of the energy of a muon produced on 
Earth to the energy $\rm E_p=m_p\,c^2\,\gamma_{in}$ of the CB's nucleons,
as they enter the SNS. We obtain a muon flux per incident CB:
\begin{eqnarray}
\rm {dn_\mu \over dx_\mu\, d\Omega}&\sim&
\rm K\,E_p^2\,{N_b\,B_\nu\over 2\,\pi\, D_L^2}\;(1+z)^4\;
f_\mu(z,\gamma_{in},\theta,x_\nu)\nonumber\\ \rm
f_\mu&=&\rm 
\;{2+c_\nu\,x_\mu\over c_\nu^3}\;e^{-c_\nu\,x_\mu}\, ,
\label{muhere}
\end{eqnarray}
with $\rm B_\nu$ and $\rm c_\nu$ as in Eq.(\ref{xtheta3})
and $\rm N_b$ the total baryon number of the CB, Eq.(\ref{NB}).
In Fig.(\ref{spectrax}c)  we show $\rm f_\mu$ as a function of
$\rm x_\mu$ at various $\theta$,
for $\rm z=1$ and $\rm\gamma_{in}=10^4$.

Very roughly characterize the efficiency of an experiment as
a step function jumping from zero to unity at $\rm E^\mu=E^\mu_{min}$.
The observable number of muons per CB and
per unit area, obtained by integration of Eq.(\ref{muhere}), then is: 
\begin{eqnarray}
\rm {dn^T_\mu[x^\mu_{min},\theta]\over d\Omega}&\sim&\rm 
{dn^T_\mu[0,0]\over d\Omega}\;G_\mu(z,\gamma_{in},\theta,x^\mu_{min})
\nonumber \\ \rm
G_\mu&=&\rm \left[{(1+z)\,b_\nu\over c_\nu}\right]^4
\;\left(1+{c_\nu\,x^\mu_{min}\over 3}\right)\;e^{-c_\nu\,x^\mu_{min}}\nonumber\\
\rm x^\mu_{min}&\equiv&\rm {E^\mu_{min}\over m_p \,\gamma_{in}}
\nonumber\\ \rm
{dn^T_\mu[0,0]\over d\Omega}
&\simeq&\rm {3.2\times 10^2\over km^2}\,{E_{CB}\over 10^{52}\; erg}\,
\left[{\gamma_{in}\over 10^4}\right]^3\,
\left[{D_L(1)\over D_L(z)}\right]^2\, .
\label{muonseen}
\end{eqnarray}

\begin{figure}
\begin{center}
\vspace*{1.0cm}
\hspace*{-1cm}
\epsfig{file=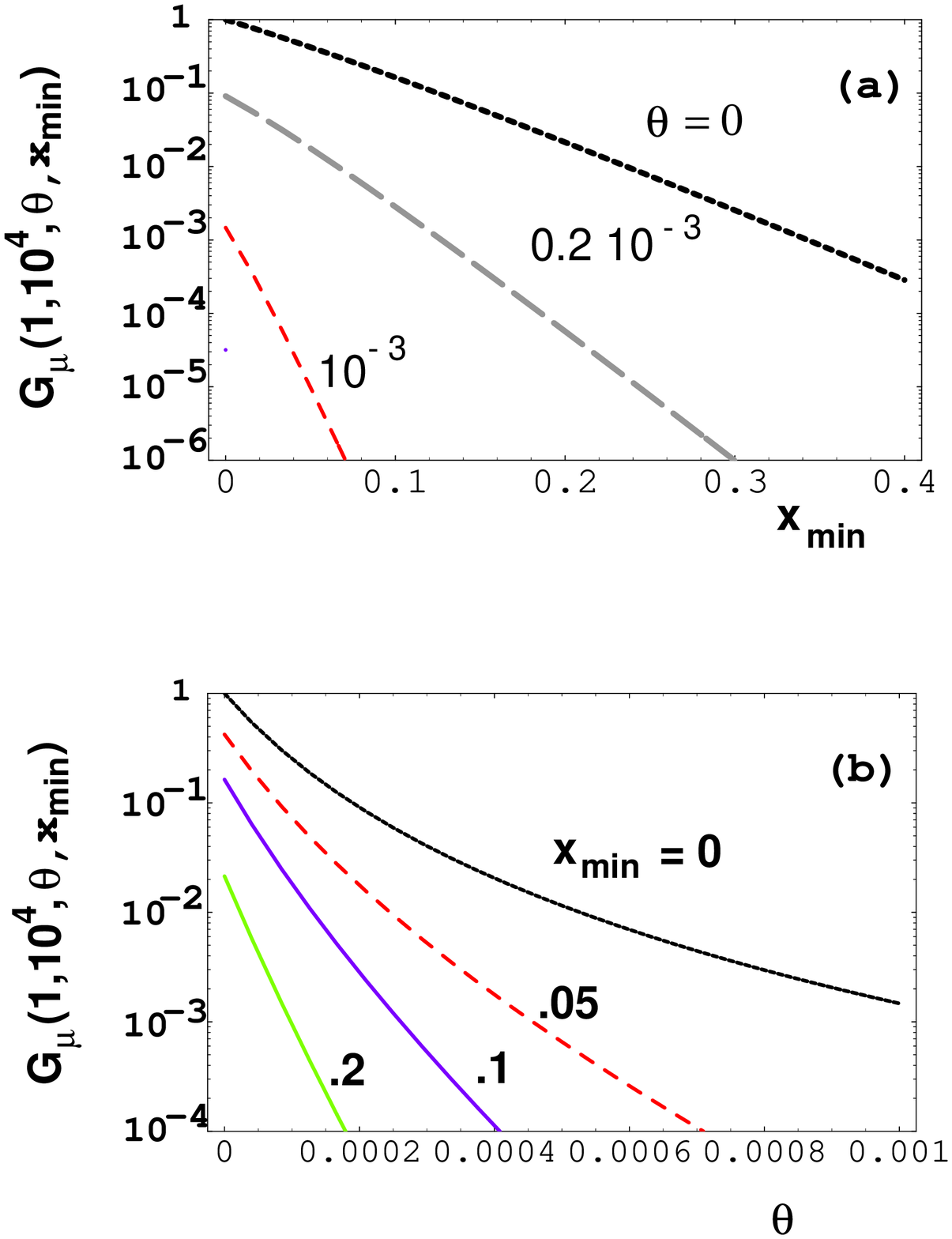,width=14cm}
\caption{The function $\rm G_\mu$ of Eq.(\ref{muonseen}), for
$\rm z=1$ and $\rm \gamma_{in}=10^4$. Top: As a function of
$\rm x_{min}$ at various fixed $\theta$. Bottom: vice versa.}
\vspace*{-0.5cm}
\label{Gmuon}
\end{center}  
\end{figure}

In Figs.(\ref{Gmuon}a,b) we show $\rm G_\mu$ as a 
function of $\rm x^\mu_{min}$ at various fixed $\theta$, and vice versa;
for $\rm z=1$ and $\rm\gamma_{in}=10^4$. The relatively large flux
$\rm dn^T_\mu[0,0]/ d\Omega$ of Eq.(\ref{muonseen}) is seen
to be very significantly reduced as $\theta$ and/or $\rm x^\mu_{min}$
 depart from zero. Once again, for
a GRB with $\rm n_{CB}$ significant pulses,
the total number of muons is $\rm n_{CB}$ times
larger than that of  Eq.(\ref{muonseen}), and neutrino oscillations
may reduce the $\nu_\mu$ flux by a factor 2 or 3.

\section{Angular apertures and observational prospects}

\begin{figure}
\begin{center}
\vspace*{1.0cm}
\hspace*{-1cm}
\epsfig{file=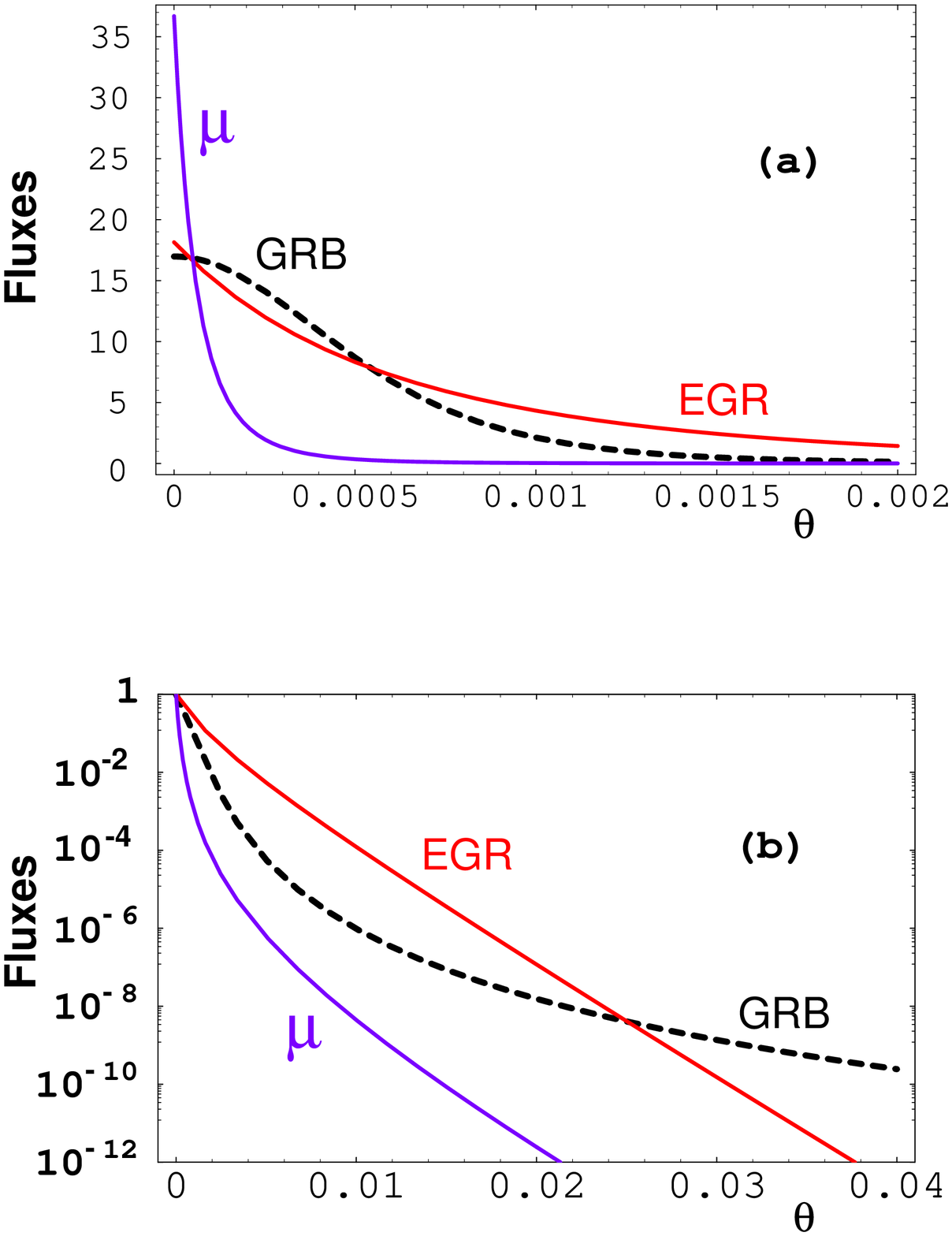,width=14cm}
\caption{Comparisons of angular distributions of
GRB photons, EGR photons and $\nu$-produced muons
in water or ice. In the upper graph, the normalizations
of the three curves are arbitrary. In the lower one,
they are all normalized to unity at $\theta=0$.}
\vspace*{-0.5cm}
\label{angdistrrs3}
\end{center}  
\end{figure}

The fluxes of $\nu$-induced muons and of $\gamma$'s
of GRB and EGR energies 
have different $\theta$ dependences, as shown
in Figs.(\ref{angdistrrs3}), where we compare the angular apertures of
the three fluxes. The absolute and relative normalizations in these
figures are arbitrary, so that the GRB results, based on
Eq.(\ref{dfdomega}), depend only on $\rm\gamma_{out}$,
chosen to be $10^3$. The EGR results, based on 
the second of Eqs.(\ref{photflux2}),
depend also on $\rm z$ (chosen at $\rm z=1$) 
and on $\rm E^\gamma_{min}$, taken here to be 50 GeV.
The $\nu$ results, also for $\rm z=1$, are based on the second
of Eqs.(\ref{muonseen}); they are for $\rm \gamma_{in}=10^4$ and
$\rm E^\mu_{min}=50$ GeV.

According to Figs.(\ref{angdistrrs3}), the EGR
beam, up to very large $\theta$, has a broader tail than the GRB beam.
In practice that means that a detector with the sensitivity to observe
the EGR flux of Eq.(\ref{photflux2}) should find a signal in temporal
and angular coincidence with a large fraction of detected GRBs.
The $\nu_\mu$-induced $\mu$ beam is about an order of magnitude
narrower than the GRB beam in angle, two orders of magnitude
in solid angle. Consequently, a detector with a sensitivity close to
that necessary to observe the $\mu$ flux of Eq.(\ref{muonseen})
would see coincidences with only about one in a hundred intense
GRB events. By ``intense'' we mean the  $\sim 1$\% of GRBs in
 the upper decade of observable fluences,
for which $\rm \theta\sim 1/\gamma_{out}$.

To ascertain the observational prospects for EGRs and $\nu$'s,
one would have to convolute our predicted fluxes
with the sensitivities of the many large-area or large-volume
$\nu$ and EGR ``telescopes''
currently planned, deployed, or under construction.
We do not have sufficiently detailed information to do so,
but a coarse look at their potential indicates that
testing the CB model will neither be trivial, nor
out of the question. The small area of past detectors with
a capability to see EGRs, such as EGRET, would preclude
the observation of the flux of Eq.(\ref{photflux2}). But the
near future looks much brighter.

\section{Timing considerations}
 
\begin{figure}
\begin{center}
\vspace*{1.0cm}
\hspace*{-1cm}
\epsfig{file=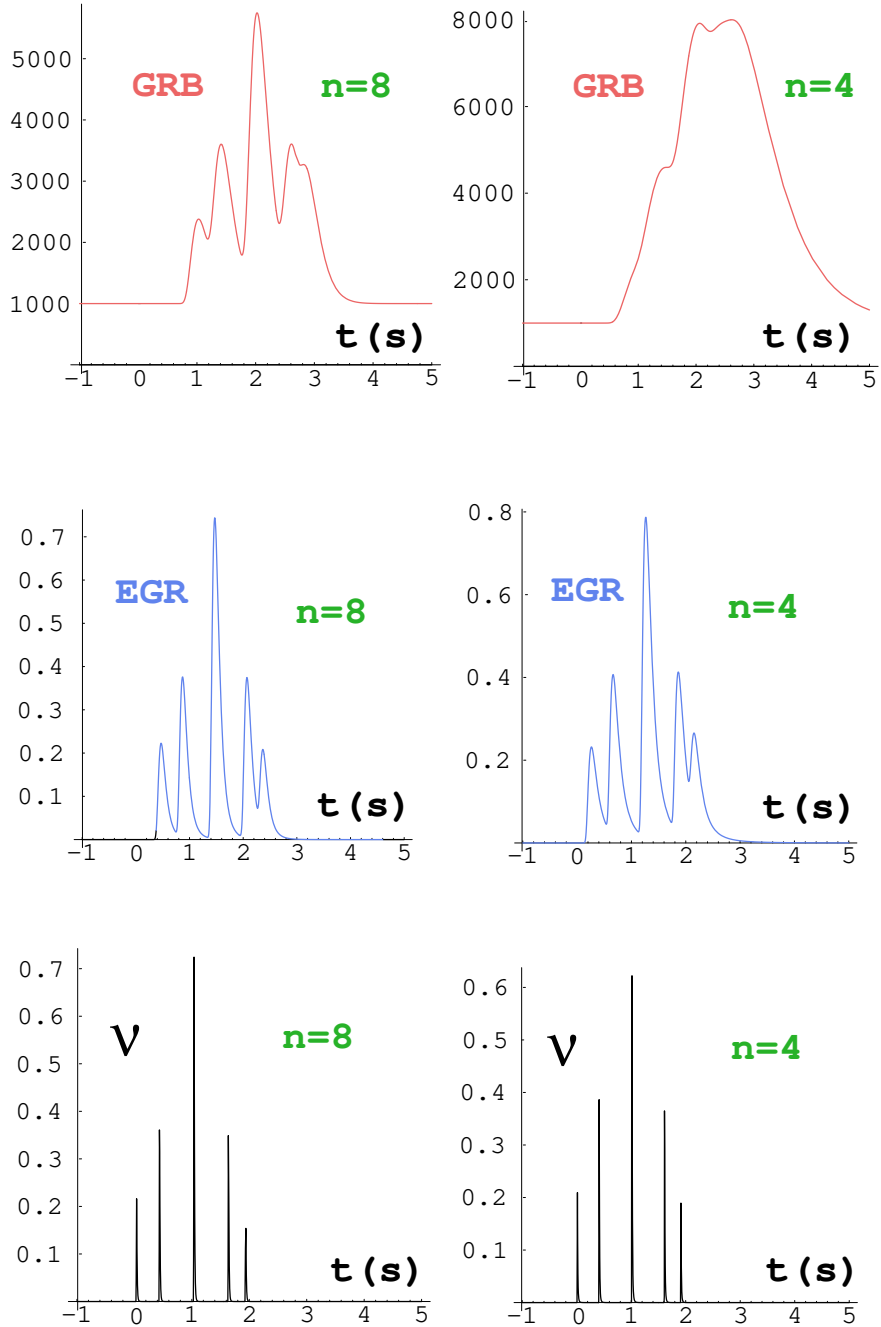,width=11.6cm}
\caption{A ``synthetic'' $\gamma$-ray burst consisting
of five CBs with $\rm\gamma_{out}$ within a factor
of 2 of $\rm\gamma_{out}=10^3$, with other parameters
at their reference values. The CBs are fired at random times in a 
2.5 s interval. The two columns are for SNS 
density indices $\rm n=8$ and 4. Top: the event seen
in the 30 keV to 1 MeV GRB domain. Middle: seen
in EGRs from $\pi^0$ decay. Bottom: the neutrino signal.}
\vspace*{-0.5cm}
\label{lightcurves}
\end{center}  
\end{figure}

In the cannonball model, each CB crossing the SNS generates
an individual $\gamma$-ray pulse in a GRB light curve. The
complementary statement need not be true: not every
observed pulse necessarily corresponds to a single CB, since
the $\gamma$ rays generated by sufficiently close CBs may
overlap. This can be seen in the two top entries in Fig.(\ref{lightcurves}),
which show the lightcurves of the same ensemble of CBs
crossing two SNSs, which differ only in their density-profile index;
in the case of the more extensive SNS ($\rm n=4$) the various
CBs blend into a single pulse.

Each CB should generate three distinct pulses: a GRB pulse,
a $\nu$ pulse and an EGR pulse. The $\nu$ and EGR pulses
are narrower in time than the GRB pulse and they precede it.
Observed with neutrinos or EGRs, then, a burst has the same
pulse structure as the GRB, but the pulses are shorter and
are precursors of the GRB pulses. These effects are
 illustrated in Fig.(\ref{lightcurves}),
where we have drawn the light curves of a single GRB in 
GRB $\gamma$-rays, in EGRs and in neutrinos. The timing sequence
of the pulses is put in by hand and their normalizations correspond
to (random) values of $\rm\gamma_{out}$ close to $10^3$. 
 The two columns of the figure correspond
to $\rm n=8$ and $\rm n=4$. Notice how the EGR pulses precede
the GRB pulses and are narrower: the EGR has a better time
``resolution''. For neutrinos, this is even more so.
It would be fascinating ---and most informative--- to ``see'' 
GRBs in two or even three complementary ways.

\section{Conclusions}

In the CB model of GRBs, illustrated in Fig.(\ref{model}), cannonballs heated
by a collision with intervening material produce GRBs by
thermal emission from a fast-cooling surface, and their electron constituency
generates GRB afterglows by bremsstrahlung
and by up-scattering of the cosmic background radiation.
The material CBs hit is an excellent ``beam-dump'',
so that nucleon--nucleon collisions
generate a very intense and collimated flux of neutrinos.
Because of absorption, the emission  of energetic
$\gamma$-rays via $\pi^0$ production and decay
is much less efficient, but by no means negligible.

\begin{figure}
\begin{center}
\vspace*{1.0cm}
\hspace*{-1cm}
\epsfig{file=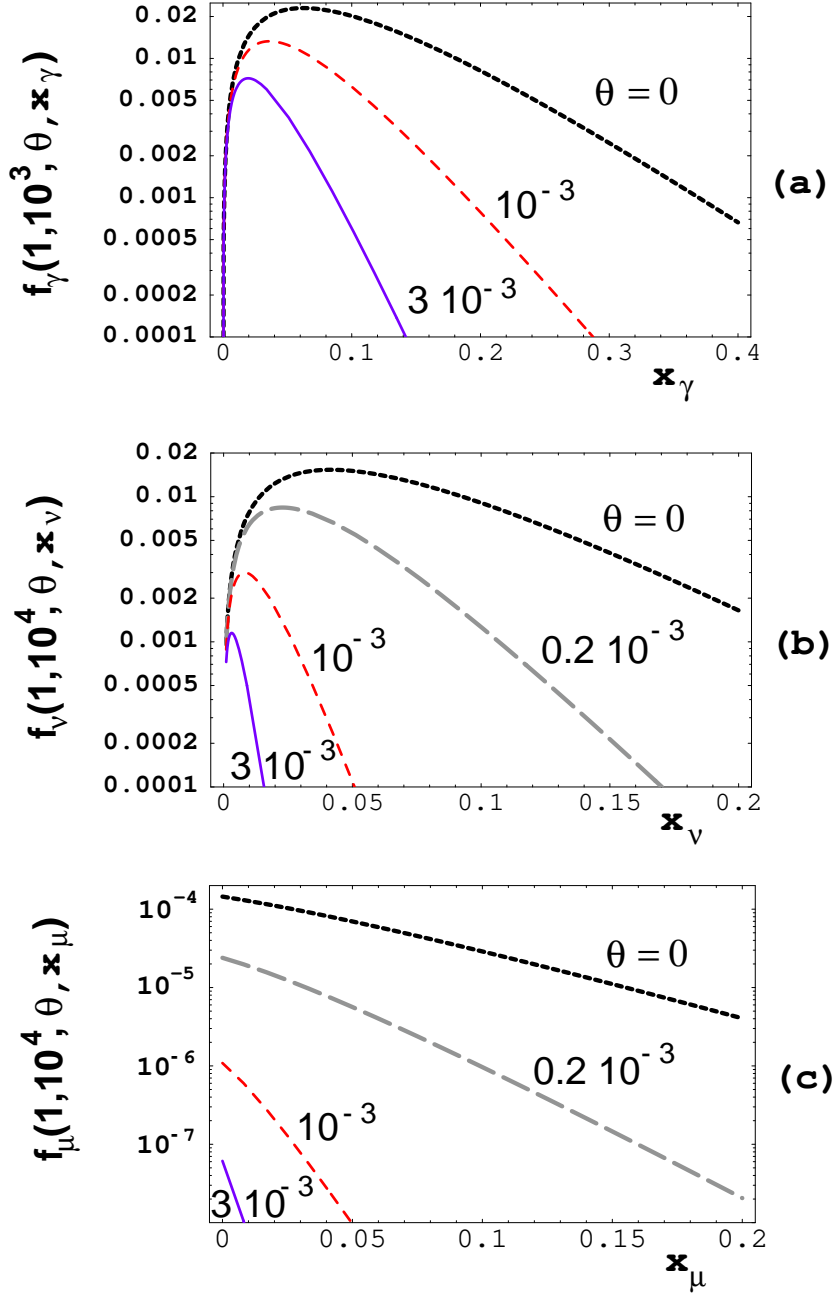,width=11 cm}
\caption{EGR, neutrino and muon fluxes, at various fixed observation
angles $\theta$, as functions of the fractional momentum of the
observed particle, at redshift unity. The functions 
$\rm f_\gamma(z,\gamma_{out},\theta,x_\gamma)$ of Eq.(\ref{photnum}),
for $\rm\gamma_{out}=10^3$, and 
$\rm f_{\nu,\mu}(z,\gamma_{out},\theta,x_{\nu,\mu})$ of 
Eqs.(\ref{nunum}, \ref{muhere}), both
for $\rm\gamma_{in}=10^4$, are depicted.}
\vspace*{-0.5cm}
\label{spectrax}
\end{center}  
\end{figure}

The $\nu$ flux has a total energy of the order 
of $10^{52}$ erg (roughly 1/3 of the total energy in a jet
of CBs,  reduced by the redshift factor).
But individual neutrinos have energies of only a few hundred GeV,
as illustrated in Fig.(\ref{spectrax}),
and their enormous flux will be hard to detect, even though it is
collimated within an angle $\sim 10^{-4}$. The detection in
coincidence with GRBs will be further hampered by the fact
that the GRB angular distribution is broader, as shown in
Fig.(\ref{angdistrrs3}).

The EGR flux carries roughly as much energy as the
GRB, that is $\rm E^{rest}_{pulse}\,\gamma_{out}\sim 10^{48}$ erg per pulse,
with $\rm E^{rest}_{pulse}$ as in Eq.(\ref{dfdomega}).
The EGR beam, as shown in Fig.(\ref{angdistrrs3}),
is somewhat broader than the GRB beam, so
that the search for coincidences should be fruitful. The typical
energies of EGRs, as illustrated in Fig.(\ref{spectrax}),
 are of tens of GeVs, and the relatively high threshold
energies of current large-area detectors should be a limiting issue,
as in the case of neutrinos.

The pulses of the GRB $\gamma$-rays should be slightly
preceded by narrower pulses of EGRs and by much
narrower pulses of $\nu$'s, as illustrated in Fig.(\ref{lightcurves}).
The CB model, as we have seen, predicts very specific properties and
relations between the GRB, EGR and $\nu$ spectra and light curves.
In this respect, as in many others, the Cannonball
Model is exceptionally falsifiable.

Gamma Ray Bursts are notoriously varied and mysterious. 
The description of their sources is clearly a multi-parameter affair.
No simplifying model, such as the cannonball model,
is going to describe in detail all of the properties of all
of these signals. Nobody in his right mind can claim
to have neatly untied the Gordian knot of the GRBs conundrum.
Yet, in moments of optimism, we believe that we have sliced it open.

\section{Acknowledgements}
I am indebted to Milla Baldo Ceolin for her excellent organization and
hospitality.

\end{document}